%% file: 2003-34n.tex
\documentclass[11pt]{article}
\usepackage[russian,english,]{babel}
\usepackage{fullpage}
\usepackage{ihep}
\usepackage{epsfig}
\usepackage{here}
\usepackage{author}
\usepackage{amsmath}
\usepackage[colorlinks]{hyperref} 

\begin{document}
\begin{titlepage}

\prepnum{2003--34}{}

\author{\Large A.~S.~Siver, ~V.~V.~Ezhela}

\title{ \Large \bf O\uppercase{n the {\boldmath$CODATA$} Recommended~Values 
of~the~Fundamental Physical Constants: V3.2(1998) \& V4.0(2002)}}


\end{titlepage}

\begin{abstractpage}[539.1.01]
\engabs
{Siver A.S., Ezhela V.V.}
{ On the CODATA Recommended Values of the Fundamental
Physical Constants: V3.2(1998) \& V4.0(2002)}

With the help of special  program package PAREVAL designed in 
{\sl Mathematica} system we reproduce values of basic fundamental 
physical constants obtained by NIST and recommended by CODATA 
for international usage since 1998. In our adjustment we use the input 
data and methods published by NIST in 1998.

It is shown, that the detected earlier inaccuracy of published by NIST 
correlations (that made the NIST 1998\&2002 results doubtful) are, 
most probably, due to inadmissible independent rounding.  

The simple estimate of the critical numbers of decimal digits in the 
independently rounded correlation coefficients is obtained. 
Further independent rounding of ``critically rounded'' correlations 
can lead to the non positive semi definite correlation matrices and 
hence is inadmissable.

 It is demonstrated by a few examples that the poor presentation of 
 the correlated random
 quantities in the scientific literature is a common bad practice and 
 is argued (once again) that the common standard for presentation  numerical values of correlated quantities in publications and sites is urgently 
and badly needed.

\rusabs
{Сивер А.С., Ежела В.В.}
{О численных значениях 
фундаментальных физических постоянных, рекомендованных CODATA версий
V3.2(1998) и V4.0(2002)}

С помощью специального пакета программ PAREVAL, разработанного в системе
{\sl Mathematica}, воспроизведены значения базовых фундаментальных 
физических постоянных NIST, рекомендованных CODATA для международного 
использования с 1998 года.  Показано, что обнаруженные ранее неточности значений
коэффициентов корреляции погрешностей констант 1998 года обусловлены, скорее всего, их недопустимым  независимым округлением. Коэффициенты корреляций
погрешностей констант 2002 года также испорчены недопустимым округлением,
и их нельзя использовать в высокоточных вычислениях.\\

  Получена простая оценка
критического числа десятичных знаков в значениях коэффициентов 
корреляции, дальнейшее независимое округление которых может приводить
к потере положительной полуопределенности матрицы корреляций.\\

На нескольких примерах показано, что некорректное представление значений
коррелированных случайных величин в публикациях --- это распространенная
вредная практика.  Декларируется необходимость разработки и использования
стандарта представления оценок коррелированных случайных величин 
в публикациях и на сайтах.  

\end{abstractpage}


\noindent

\vspace*{-0.5cm}
\section{Motivation}

Fundamental physical constants (FPC) are the basic entities in pure and
applied natural sciences and in technology. Thanks to efforts of many national
metrology institutions, \href{http://www.nist.gov/}{NIST}\footnote{National Institute of Standards and Technology(USA)~\cite{NIST}.}~ systematists, and international coordination
by \href{http://www.codata.org}{CODATA}\footnote{Committee on Data 
for 
Science and Technology \cite{CODATA}.}~ we {\it seem} to have a more or less reliable procedures to monitor the development of the unified system of constants, their  periodically adjusted numerical values, uncertainties, and correlations.  

We use ``{\it seem}'' because in spite of many national and international documents on the rules and standards on the statistical and experimental data presentations in the official publications\footnote{We failed, however,
to find any official documents
standardising the procedures of rounding average values, standard uncertainties,
and their correlations of the {\bf jointly measured or evaluated (adjusted)} quantities.} ~there is a sharp contradiction between rules/standards and reality of scientific data exchanges in 
the past and modern scientific communication media.

An example of such contradictions is the lack of attention from producers (NIST) and overseers (CODATA) of the evaluated FPC data to 
the quality of the final data presentations in the official publications on the paper and even in electronic forms: data presented are incomplete
and inaccurate  as we will show further.

The other example of the mentioned contradiction is the ignorance of correlations in  all respectable information resources: handbooks, textbooks, monographs, reviews,
and scientific software packages
that have reprinted samples of the recommended FPC-1998 (see, for 
example,~\cite{Lide:2001}
-- \cite{MAXpack}).

\
Since the release of FPC-1998 the scientific community obtained real access to the correlations of the FPC uncertainties and one can see that correlations between uncertainties of some universal constants are too ``strong'' to be ignored in high accuracy calculations. But unfortunately the correlation
matrix presented on the NIST/CODATA site 
and reproduced partly in the publication \cite{MohrTaylor} (see, Table XXV
on the page 453) is non positive semi-definite in contradiction with definition of the correlation matrix. Looking at the NIST correlation coefficients one
can see that they are rounded off too ``tightly.'' Format of the numbers shows
that they were rounded uniformly and independently (e.g. ignoring the 
crucial constraints that any covariance and correlation matrix must respect).

\begin{table}[h]
\begin{center}
\caption{Sample of a few NIST/CODATA:1998 recommended constants.}
\vspace*{1mm}
\begin{tabular}{|l|c|l|cccc|}
\hline
 & & & \multicolumn{4}{|c|}{}\\[-3mm]
FPC name & Simbol [units] & Value (uncertainty)$\times$scale& \multicolumn{4}{|c|}{Correlations}\\[1mm]
\hline
& & & & & &\\[-4mm]
Elementary charge & $e \quad \quad [{\rm C}]$ & $ 1.602\, 176\, 462(63)\times10^{-19}$
& $e$ & $h$ & $m_e$ & $m_p$\\ \cline{4-7}
& & & & & &\\[-4mm]
 Planck constant & $h \quad \quad [{\rm J} \, {\rm s}]$ & $6.626\, 068\, 76(52) \times 10^{-34}$ & 0.999 &  & &\\ 
Electron mass & $m_e \quad [{\rm kg}]$ & $9.109\, 381\, 88(72) \times 10^{-31}$
&0.990 &0.996 & &\\  
Proton mass & $m_p \quad [{\rm kg}]$ & $1.672\, 621\, 58(13) \times 10^{-27}$
&0.990 &0.995 &1.000 &\\ [1mm] 
\hline
\end{tabular}
\end{center}
\vspace*{-8mm}
\end{table}

\vspace*{3mm}

From the textbooks on  numerical calculations it is known that the rounding of the correlated values is a subject of special treatment (see for example
\cite{Kalitkin}, page 499). The average values, standard uncertainties, 
and correlation coefficients could not be rounded off independently. 

Independent rounding
may lead to catastrophic changes in the connection of averages, standard uncertainties, and scatter ellipsoid: average values may get out of scatter
ellipsoid, scatter ellipsoid may turn to hyperboloid after independent 
rounding off the correlation coefficients.

 With the lack of discussions of the rounding correlated
quantities in the NIST/CODATA publications, we interpret this as the result of independent rounding that destroy catastrophically the system of adjusted
values. An example of such catastrophe with FPC-1998 recommended for ``public usage'' is the negative variance for the Rydberg constant calculated from the equation 
$R_{\infty}=\alpha^2 m_{\rm e} c /{2 h}$ \cite{EL-2}. If this confusion did not caused by a misprint in signs of some correlation coefficient it is 
most probably the ``inconsistency induced by rounding off'' the adjusted FPC-1998. 
 
If we are right in our account that corruption of the true data was due to
independent rounding then we should clarify the influence of independent rounding of the average values and dispersions. To test this a
 reproduction of the whole adjustments procedure used by NIST experts
is needed. 

The main goal of this work is to reproduce the NIST results using their data
and methods as they presented in the detailed publication \cite{MohrTaylor}
and on the NIST site and then to work out the proper way of correlated data presentation and exchange. 

 The other goal of our work is to draw  attention (once again) of the 
physics community to the problem with standardization of 
the statistical (experimental) data presentation in the modern 
scientific communication media.

The rest  of the paper is organized as follows. In the
next section we present a few variants of our adjustment in comparison
with corresponding NIST/CODATA results.  The last section summarizes
 outputs from our exercises and our vision how to improve  the 
 situation. 

\section{On the NIST FPC Adjustments Technology}

In our analyses we tried to be as close as possible to the NIST 
adjustments strategy. Fortunately more or less complete overview of the  NIST:1998 adjustment procedure including detailed presentation of the experimental
data, theoretical models and formulae, the FPC
evaluation strategy description, and explanation of the specific aspects of the calculations were published in~\cite{MohrTaylor}. 
To test the \href{http://ts.nist.gov/traceability/nist%20traceability%20policy-external.htm}
{\bf traceability} of the NIST results through their public information
resources we have attempted to reproduce NIST FPC values on the basis of data, formulae and instructions  from publication \cite{MohrTaylor} and 
\href{http://physics.nist.gov/cuu/Constants/index.html}{NIST} site only.\\
\begin{table}[ht]
\caption{Basic adjusted constants.}
\label{constadj}
\include{constable}
\end{table}

 There are 107 principal input data expressed as functions
of 57 adjusted constants (variables) via the set of  observational equations.
These 57 variables are subdivided in two classes: constants (see {\bf Table~\ref{constadj}})
 and corrections (see {\bf Table~\ref{corradj}}) 
 
\begin{table}[ht]
\caption{Basic adjusted corrections.}
\include{deltable}
\label{corradj} 
\end{table}

Numerical values in the {\bf Tables \ref{constadj}, \ref{corradj}} 
are presented in the standard concise form $X(Y)\times10^{Z}\,{\rm u}$ 
(where $X$ --- average value,
$Y$ --- standard uncertainty of $X$ referred to the last digits of the quoted
value, 
u --- unit)
 and are expressed in SI.
The symbols of adjusted constants are as follows:
 $R_{\infty}$ (Rydberg constant); $\alpha$ (fine structure constant);
$h$ (Plank constant); $m_e/m_{\mu}$ (electron-muon mass ratio); 
 $A_r(X)$ ( mass of the particle $X$ in atomic units), where $X$ denotes symbol
of the particle, such as electron or alpha-particle; $\mu_X/\mu_Y$ ($X$, $Y$
magnetic moment ratio); $R$ --- molar gas constant;
 $\rm xu(X\:K\alpha_1)$ ---  x-ray unit of the X atom;
 $\rm d_{220}(X)$ --- \{220\} lattice spacing of the different (X) silicon mono-crystal ($\rm d_{220}$);
$R_p$ and $R_d$ are the bound-state proton and deuteron rms charge radii.


There are also 28 adjusted variables (see {\bf Table~\ref{corradj}}) 
that are not fundamental at all, but
were introduced in order to decrease the theoretical uncertainty of the several
observational equations. They are: 

$\delta_{\rm N}(n,\rm{L},2\cdot j)$ --- additive
correction to $n \rm{L}_j$ energy level of the hydrogen (N=H) and deuterium
(N=D); 

$\delta_{e}$, $\delta_{\mu}$ --- additive correction to electron and muon magnetic moment anomaly, 

$\delta_{Mu}$ --- additive correction to 
 hyperfine splitting of the muonium basic bound state. \\

Unfortunately the set of final values released so far by NIST/CODATA:1998  does not contain estimates for large part of the 57 adjusted variables 
(marked as ``not given'' in the {\bf Tables} \ref{constadj}, \ref{corradj} 
in cases
when we failed to find corresponding output value on the site 
{CODATA Fundamental Physical Constants. Version 3.2 Release date: 1 October 2003.} \cite{versions}). Hence our comparison will be incomplete to this extent. 
\newpage

The other problem that makes the straightforward reproduction
of the NIST estimates impossible is the corrupted presentation of the input correlation
sub-matrix (see \cite{MohrTaylor}, page 434, Table XIV.A.2)
of the data related to the Rydberg constant. It is non positive
definite. We interpret this confusion as the result of unjustified 
independent rounding of correlation coefficients (motivated only to be 
convenient for publication on the paper).  Fortunately
the correlation sub-matrix for the other data presented in \cite{MohrTaylor}
(see page 436, Table XIV.B.2 ) is positive definite. 


In spite of the incomplete presentation of the adjustment
results and corrupted input correlation data we decided to clarify to what extent the ignorance of the input correlations in part or totally will modify the output constants, supposing that average values of the constants are correct as well as their standard uncertainties.

The strategy of comparisons is as follows. First of all we have convinced that the NIST method used to obtain values of adjustable variables is indeed the method to find stationary points of the ``linearized'' $\chi^2$. 
If the input covariance matrix is positive definite then the obtained solution will be a minimum and adjustment will be stable if we turn lucky to get into vicinity of a global minimum.
 If the input covariance matrix is non positive definite then
the task of finding stationary points could be solved for non-degenerate weight matrices but in this case the task has no connection with least squares method of constants estimation.

 Nevertheless it is interesting to compare values of adjusted variables that
 can be obtained by NIST method in our PAREVAL package (from the NIST starting point in the adjusted variables space and with  
 rounded by NIST input covariance matrix \cite{MohrTaylor}) with  
 supposedly true NIST:1998  values.
\vspace*{5mm}

\subsection{Calculations with input correlator published by NIST } 
\vspace*{2mm}
  Corresponding results are presented in  {\bf Table \ref{t-1-1}}, where 
  in the fifth column 
  the normalized difference
  of values of third $(X_3)$ and fourth $(X_4)$ columns is defined as follows
  \vspace*{5mm}
  $$ \Delta = \frac {X_{3}-X_{4}} {\sqrt{\sigma^2(X_{3})+\sigma^2(X_{4})}}$$
  \vspace*{5mm}
 
 In our adjustment we use close to the NIST 
starting point (used for finding zero of the gradient 
of $\chi^2$ in step-by-step method) 
in the adjusted variables space and ``weights'' were constructed from  
 rounded by NIST input ``correlation'' matrix \cite{MohrTaylor}. 
 
 As it can
 be seen from
 the fifth column of the  {\bf Table \ref{t-1-1}}, the
 shifts in average values are in general well inside the ranges defined by the quadratically combined ``uncertainties''. In all tables, as a rule, we save in average values and corresponding uncertainties
one more digit to show that some values are not reproduced exactly when rounding
independently even if
they are well inside the uncertainties ($|\Delta| < 0.1$).

\begin{table}[p] 
\caption{\small Comparison of the true NIST/CODATA:1998 recommended
 values (third column) with corresponding values (fourth column) that
 have been obtained at IHEP by the NIST method in our PAREVAL package.}
\label{t-1-1}
\vspace*{0.2cm}
\include{table1-1}
\vspace*{5mm}

\end{table}

\begin{table}[p]

\par
 As we have no  NIST/CODATA:1998  final values of the adjusted corrections we compare our results with the input values which we use (follow NIST) as observational equation like $\delta = 0.0 \dots 0(u(\delta))$
(see {\bf Table \ref{t-1-3}}).
We see that shifts in average values for all corrections are also well 
 inside the intervals defined by the input estimates of theoretical 
 systematic uncertainties.  
 \vspace*{5mm}
 
\caption{\small Comparison of the input NIST/CODATA:1998 
 values (third column) with corresponding values (fourth column) that
 obtained at IHEP by the NIST method in our PAREVAL package.}
\include{table1-3}

\label{t-1-3}
\end{table}

\begin{table}[p]

\subsection{Calculations with identity input correlation matrix}

We learn from  Tables \ref{t-1-1} and \ref{t-1-3} that rounded input 
correlation matrix leads to slightly shifted values of the basic 
constants and corrections that are well inside the uncertainties. 
To clarify the importance of correlations further and to test 
our package we produce adjustment ignoring input correlations. 
\vspace*{0.8cm}

\caption{
\small Comparison of the true NIST/CODATA:1998 recommended
 values (third column) with corresponding values (fourth column) that
 obtained at IHEP by the NIST method in our PAREVAL package. In this adjustment
 we use the same starting point  as in previous run and weights were constructed completely ignoring input correlations. We see (from the fifth column) that
 shifts in average values are in general inside the quadratically combined ``uncertainties''.}

\include{table21-1}
\label{t-21-1}
\end{table}

\begin{table}[p]
\caption{Comparison of the obtained values of additive corrections
with input estimates of the theoretical systematic uncertainties. }
\vspace*{5mm}
\include{table22-3}

\label{t-22-3}
\end{table}

\begin{table}[p]
Now we can compare values of constants and corrections  obtained with rounded
correlator and without correlations (see {\bf Tables \ref{t-3-1}, \ref{t-3-2}}). 
\vspace*{0.8cm}

\caption{Comparison of constant estimates obtained with NIST
published input correlations and with identity matrix as input 
correlation matrix}
\include{table3-1-1}
\label{t-3-1}
\end{table}

\begin{table}[p]
\caption{Comparison of correction estimates obtained with NIST
published input correlations and with identity matrix as input 
correlation matrix.}
\include{table3-2-1}

\label{t-3-2}
\end{table}

\newpage
Finally we have checked if the procedure to find minimum of the corresponding $\chi^2$ without linearization of the observational equations  used 
gives the
same result. We have used the built-in {\it Mathematica} module to find minimum starting from the point in the adjusted variables space where the values  of all 29 constants were taken as recommended by NIST/CODATA:1998 and the
 values of the rest 28 $\delta$-corrections were taken
as $0.001\times u(\delta_{theor})$.

 Corresponding value of $\chi^2/ndf=0.84$, i.e. the same as obtained by 
 using NIST procedure with linearization.  Average values of the 
 constants and corrections obtained by two different methods and 
 different programs are practically the same (the maximal normalized 
 difference is $|\Delta|_{max} 
 \sim 10^{-21}$).
 
Our results lead us to the conclusion  that NIST 
experts have used input correlation matrix close to that of presented in
their published report.
  The fact
 that output correlation matrix is also non positive semi-definite does  not allow one to exclude possibility that this non positive semi-definiteness could
 be induced by the non positive definiteness of the input matrix. This point
 remains to be clarified. 
   
 Let us compare a sample of our output correlation coefficients  with 
 corresponding NIST values.
 \begin{table}[h]
 \caption{The values in bold are extracted from the NIST site whereas the  
 values placed under main diagonal are our values obtained with no input
 correlations.}
 \include{cor-spec}
 \label{correl}
 \end{table}
 
As shown in the {\bf Table \ref{correl}} some of our correlation 
coefficients differ significantly from those of published by NIST.

Now we  proceed to the general comments on the practice in the scientific literature and on the sites, where the correlated 
estimates of the random quantities jointly measured or evaluated 
are presented. 

 Often authors of the original
papers appear to ignore correlations at all or to present them in an 
incomplete manner. So, it is hard to 
understand what type of uncertainties the quoted correlation matrix 
is referred to: statistical, systematic or total. It is dangerous 
(or even inadmissable) to use such incomplete data in further 
analyses and especially in the theory tests. Sometimes this 
incompleteness is caused by the too firm editors 
and publishers requirements. On the other hand there are also 
experimental mistakes 
(see examples of such situation in nuclear physics and technology
 \cite{Badikov:1998}).

\subsection{On the rounding off the correlated estimates}

As it is well known the covariance matrix for the jointly estimated
statistical quantities is by definition a positive semi-definite 
Hermitian matrix. It  has real and nonnegative eigenvalues.
The non degenerate correlation matrix is positive definite by 
definition. 

Unfortunately some authors 
publish the ``correlation matrix'' with no final check up 
this crucial property of the correlation matrix.  
In majority cases it happens  
under
the pressure of the limited publication space. So, authors are forced 
to present 
rounded correlation coefficients making this in an inadmissable manner.
The rounding off the correlation coefficients are produced independently,
saving symmetry of the matrix but ignoring such crucial 
properties as positive definiteness and positive 
semi-definiteness.

To be specific we quote a few examples from different subject fields. 
The first most
striking example is yet discussed concerning the NIST publications 
on the adjusted fundamental constants, including the NIST site. The 
published
version of the input correlation sub-matrix used to construct the 
weight matrix in their version of least squares method (LSM) is non
positive definite, (see \cite{MohrTaylor}, page 434, Table XIV.A.2),
it has two negative eigenvalues. Also the published version of the 
correlation sub-matrix between uncertainties in the recommended 
values of a sample of fundamental constants (see \cite{MohrTaylor}, 
page 453, Table XXV) is non positive semi-definite. We have convinced 
that it is because of poor accuracy of the presentation caused by 
unjustified uniform independent rounding of the correlation coefficients.
Unfortunately the same way of presentation is used by NIST and 
approved by CODATA on their sites.  

The other example is the publication of the CLEO collaboration
on the high precision measurements of the $\tau$-lepton decay 
branching ratios \cite{Anastassov:1996tc}. The final version of the
correlation matrix presented in the Erratum is as follows:
\begin{table}[h]
\begin{center}
\caption{Correlation coefficients between branching fraction measurements
of some $\tau$-lepton decays from \cite{Anastassov:1996tc}
.}
\vspace*{2mm}
\begin{tabular}{c|ccccc}
 {\boldmath \large$\tau$}  &$ B_e $ & $ B_{\mu} $ & $ B_h  $ & $ B_{\mu}/B_e $ &  $ B_h/B_e $\\[1mm] \hline
 & & & &       & \\[-4mm]
$ B_e $        & 1.00 & 0.50 & 0.48 &$-$0.42 &$-$0.39 \\   
$ B_{\mu}$     &      & 1.00 & 0.50 &\phantom{$-$}0.58 &\phantom{$-$}0.08 \\
$ B_h    $     &      &      & 1.00 &\phantom{$-$}0.07 &\phantom{$-$}0.63 \\ 
$ B_{\mu}/B_e$ &      &      &      &\phantom{$-$}1.00 &\phantom{$-$}0.45 \\  
$ B_h/B_e$ &      &      &      &        &\phantom{$-$}1.00 \\  
\end{tabular}
\end{center}   
\label{cleotab}
\end{table}

\noindent
The corresponding eigenvalues are as follows:
$${2.173\,46,\quad 1.781\,87,\quad 1.054\,97,\quad -0.007\,491\,53,
\quad -0.002\,803\,4}$$
 This confusion could be due to improper rounding, but we failed to 
 show this by playing with numbers (un-rounding).  
 The problem seems to be deeper and hence the CLEO data are questionable. 
 These should be used with great caution in theory tests and in 
 derivations of ``world averaged'' $\tau$-lepton branching ratios.

As we already mentioned, the proper rounding procedure for the jointly 
measured or estimated quantities (average vector components, corresponding vector of their standard uncertainties, and correlation matrix) is the 
subject of special treatment. So it will be presented elsewhere if we will not found relevant papers published.

In the next section we  
construct a simple but important estimate of the threshold accuracy 
of the correlation coefficients that should not be violated 
while uniform independent rounding of correlation matrix elements.
\section{On the numerical presentations of correlated quantities  
in computer readable files and in publications}

Here we derive a simple sufficient estimate on the accuracy of a safely
independent and uniform rounding the correlation matrix elements off.

Let $A_{ij}$ be the $n \times n$ correlation matrix. It is real, 
symmetric, positive definite, and has matrix elements bounded 
as follows 
$$A_{ii}= 1 \quad {\rm for\, all} \quad i=1,\dots,n \quad {\rm and} 
\quad  |A_{i \ne j}| < 1.0.$$

\noindent
Let $B_{ij}$ be the ``rounder'' matrix, such that if it is added 
to the matrix $A_{ij}$ the obtained matrix $G_{ij} = A_{ij} + B_{ij}$ 
will be real, symmetric, positive definite and all 
$|G_{i\ne j}| < 1$ are decimal numbers with $k$ digits 
after the decimal point.

It is easy to see that matrix $B_{ij}$ should have the following properties:
$$B_{ii}= 0 \quad {\rm for\, all} \quad i=1,\dots,n \quad {\rm and} 
\quad  |B_{i \ne j}| \le  5.0 \times 10^{-k-1}.$$

Let further $\alpha_1 \le \dots \le \alpha_n$, 
$\beta_1 \le \dots \le \beta_n$, and 
$\gamma_1 \le \dots \le \gamma_n$ be the ordered sets  of eigenvalues 
of the matrices  $A_{ij}$,  $B_{ij}$, and  $G_{ij}$ correspondingly.
Then from the Weil's theorem for any $l = 1,\dots,n$ we have 
the following  inequalities \cite{HornJohnson},\cite{Wilkinson}:

$$\alpha_l+\beta_1 \le \gamma_l \le \alpha_l + \beta_n.$$ 

From the Gershgorin's theorem on the distributions of the eigenvalues of
the Hermitian matrices \cite{HornJohnson}  it follows that 
$$\beta_1 \ge -(n-1)\cdot 5\cdot 10^{-(k+1)} = 
-{\frac{(n-1)} {2}}\cdot10^{-k}$$
 and hence to have the matrix $G_{ij}$  as positive semi 
definite matrix it is sufficient to demand 
$$0 \le \alpha_1 - {(n-1)\over 2}\cdot10^{-k} \le \gamma_1.$$

From the left inequality we have the final estimate for the 
threshold accuracy index for safely uniform independent rounding (SUIR) of the 
positive definite correlation matrix $A_{ij}$ with minimal 
eigenvalue $\alpha_{min}$
{\boldmath \large 
\begin{equation}
{ k \ge K_{\rm SUIR}^{th} = \left \lceil \log_{10}\left({{n-1} \over 
{2 \cdot \alpha_{min}}}\right ) \right \rceil}.
\label{K_SIR}
\end{equation}    
}  
  
{\bf NOTE.} According to the Weil's theorem  any uniform rounding 
 the off-diagonal matrix elements of the positive semi-definite covariance matrix is forbidden.

 Indeed, as rounder
matrix is traceless Hermitian matrix, it obliged to have the 
negative minimal eigenvalue. Furthermore from the left inequality
of the Weil's theorem statement it follows that any rounding
could lead to the matrix with negative minimal eigenvalue.

This note shows that the special rounding strategy should be 
developed\footnote{ We realise that, most probably, such a strategy 
was developed already somewhere, but unfortunately is deeply hidden in 
the national and international metrology instructions.
Some relevant information see, for 
example, in the review \cite{Badikov:1998}, where the analogous concerns
are expressed.}~ for such covariance matrices as well as for the 
badly conditioned covariance matrices.

Now we can make a further comments on the values of the FPC correlation 
coefficients. In the Table \ref{correl},  to simplify comparison with NIST data, our output correlation 
coefficients for basic FPC are 
presented with independent rounding to three significant digits. In fact we have minimal 
eigenvalue for our output $57\times57$ correlation matrix 
 $$\boldmath{\alpha_{min} = 1.290\,020\,861\,518\,11 \times 10^{-6}}$$ 
and
from the expression (\ref{K_SIR}) for the critical accuracy it follows  
that {\it at least 8 digits 
after decimal point should be saved, when rounding uniformly and 
independently,  to preserve positive  
 definiteness of the output correlation matrix for basic sample of FPC.}

\begin{table}[h]
\begin{center}
\caption{Comparison of a few CODATA:1998 and CODATA:2002 recommended constants.}
\vspace*{1mm}
\begin{tabular}{|l|c|l|cccc|}
\hline
 & & & \multicolumn{4}{|c|}{}\\[-3mm]
{\bf \large CODATA:1998 } & Simbol [units] & Value (uncertainty)$\times$scale& \multicolumn{4}{|c|}{Correlations}\\[1mm]
\hline
& & & & & &\\[-4mm]
Elementary charge & $e \quad \quad [{\rm C}]$ & $ 1.602\, 176\, 462(63)\times10^{-19}$
& $e$ & $h$ & $m_e$ & $m_p$\\ \cline{4-7}
& & & & & &\\[-4mm]
 Planck constant & $h \quad \quad [{\rm J} \, {\rm s}]$ & 
 $6.626\, 068\, 76(52) \times 10^{-34}$ & 0.999 &  & &\\ 
Electron mass & $m_e \quad [{\rm kg}]$ & $9.109\, 381\, 88(72) \times 10^{-31}$
&0.990 &0.996 & &\\  
Proton mass & $m_p \quad [{\rm kg}]$ & $1.672\, 621\, 58(13) \times 10^{-27}$
&0.990 &0.995 &1.000 &\\ [1mm] 
\hline \hline
 & & & \multicolumn{4}{|c|}{}\\[-3mm]
{\bf \large CODATA:2002 } & Simbol [units] & Value (uncertainty)$\times$scale& \multicolumn{4}{|c|}{Correlations}\\[1mm]
\hline
& & & & & &\\[-3mm]
Elementary charge & $e \quad \quad [{\rm C}]$ & $ 1.602\, 176\, 53(14)\times10^{-19}$
& $e$ & $h$ & $m_e$ & $m_p$\\ \cline{4-7}
& & & & & &\\[-3mm]
 Planck constant & $h \quad \quad [{\rm J} \, {\rm s}]$ & 
 $6.626\, 0693(11) \times 10^{-34}$ & 1.000 &  & &\\ 
Electron mass & $m_e \quad [{\rm kg}]$ & $9.109\, 3826(16) \times 10^{-31}$
&0.998 &0.999 & &\\  
Proton mass & $m_p \quad [{\rm kg}]$ & $1.672\, 621 \,71(29) \times 10^{-27}$
&0.998 &0.999 &1.000 &\\ [1mm] 
\hline
\end{tabular}
\end{center}
\end{table}

\noindent
Eigenvalues of these correlation sub-matrices are as follows:
$$ CODATA:1998 \quad \{3.985,\, 0.0150769,\, 0.00536526,\, -0.000617335\};$$
$$ CODATA:2002 \quad \{3.997,\, 0.00315831,\, -0.000158432,\,-2.83681\times10^{-16}\}
. $$

\noindent   
Both matrix are non positive semi-definite, the 2002 sub-matrix is degenerate
as it is seen from the table above.

Another concern is the accuracy of data presentation on average values and
standard uncertainties. As a rule 2002--uncertainties are more than two times larger than corresponding 1998-numbers (see the table above).

 It is to some
extent unexpected as the NIST bibliography database on FPC contains 528 additional
to 1998 database entries
classified as ``experimental'' and ``original research'' dated between
1999 and 2002 inclusively. Unfortunately the reasons of these enlargements of the standard
uncertainties compared to the V3.2(1998)--release did not commented in the V4.0(2002)--release notification.\\
\newpage

\section{Conclusions}

In this section we summarize the main results obtained and the discussions
presented above.

\begin{itemize}
\item
The {\sl Mathematica}  package PAREVAL was created  and applied for 
adjustments of FPC. It reproduces the
 NIST/CODATA adjustment technology from their data and methods 
 (as of 1998). It 
is composed of a few modules to maintain: the library of theoretical models,
the experimental data compilation; to perform FPC evaluations, and 
for results presentation. The detailed package description as well as 
the address for access will be presented elsewhere \cite{Siver:2003}.
       
\item
With the help of PAREVAL it is shown, that the ``CODATA recommended 
values of the fundamental 
physical constants: 1998''  V3.2 and 
``CODATA recommended 
values of the fundamental 
physical constants: 2002''  V4.0 
are questionable in, at least, the values of 
published correlation coefficients released in the NIST/CODATA sites. 
Most probably data were corrupted by unjustified rounding up the 
output values.\\

  It is argued that the released so far correlation 
coefficients are useful only to show the sizes of correlations but 
should not be used in the real calculations of high precision 
observables. It will be extremely useful if the released for the first time
ASCII file be accompanied with easy computer readable files with 
the compact standardized names, units, average values, standard 
uncertainties and correlations     
presented as accurate as possible, without unjustified rounding up.

\item
  The simple estimate for the threshold accuracy 
sufficient for safely uniform independent rounding of the positive 
definite correlation matrix is constructed. This estimate can be 
used to trigger the data corrupted by unjustified rounding. 
\item
It is argued that it is an urgent need to create a common standard 
strategy of rounding interrelated quantities and common standard
data structures to store and exchange the correlated data in the 
computation media. These standards should be freely 
available for science, education and technology practitioners.

 It seems that it is a real challenge to IT 
professionals to construct a flexible and tractable technology 
to handle large samples of correlated data which will preserve
all global properties and interconnections of the principal 
components of the stricture in all data transformations and 
exchanges.
\end{itemize}

%
{\bf Acknowledgements}\\
\vspace*{-2mm}

This work was supported in part by the Russian Foundation for Basic Research
under grant RFBR-01-07-90392. 

We gratefully acknowledge the encouraging 
remarks from R.M.~Barnett, P.~Langacker, and T.G.~Trippe.

 We thank our 
colleagues V.B.~Anikeev, Yu.V.~Kuyanov, V.N.~Larin, S.B.~Lugovsky, 
K.S.~Lugovsky, A.D.~Ryabov, N.P.~Tkachenko, and O.V.~Zenin  for the 
fruitful discussions and especially to I.I.~Degtyarev who informed 
us on the similar situation with treatment of correlated data in 
nuclear metrology community and gave us relevant references. We 
thank S.A.~Badikov for reading manuscript and comments.


\enterdate{December 4, 2003,\\ Revised December 17, 2003.~}

\end{document}

%% file: constable.tex
\begin{center}
{\footnotesize
\begin{tabular}{|r|l|c|c|}
\hline 
 & & & \\[-3mm]
& Symbol  &  FPC-1998 values V3.2& Connections\\ 
\hline
   & & &  \\[-3mm]
1  & $ R_{\infty}          $   & 
$10\, 973\, 731.568\, 549(83)\,{\rm m^{-1}}$ & $R_{\infty}=\alpha^2 m_{\rm e} c /{2h}$ \\ \hline
& & & \\[-3mm]
2  & $ A_r(e)              $ &
$5.485 \,799\, 110(12) \times 10^{-4}\, u $ & \\ 
3  & $ A_r(p)              $ & 
$1.007\, 276\, 466\, 88(13)\, u$  & \\
4  & $ A_r(n)              $ &
$1.008 \,664\, 915\, 78(55)\, u $    & \\ 
5  & $ A_r(d)              $ &  
$ 2.013 \,553\, 212\, 71(35)\, u $   &\\
6  & $ A_r(h)              $ &
 $ 3.014\, 932\, 234\, 69(86)\, u $ & \\
7  & $ A_r(\alpha)         $  &
 $4.001\, 506\, 1747(10) \,u  $    & \\ \hline
 & & & \\[-3mm]
8  & $ \alpha              $ & 
$7.297\, 352\, 533(27) \times 10^{-3}$ & $\alpha = e^2/{2 \epsilon_0 h c}$ \\ \hline
& & & \\[-3mm]
9  & $ \mu_{e^{-}}/\mu_p   $&
$ -658.210\, 6875(66) $   & \\
10 & $ \mu_d/\mu_e^{-}     $ &
$-4.664\, 345\, 537(50)\times 10^{-4}$ & \\
11 & $ \mu_{e^{-}}/\mu_p'  $ &
$ -658.227 \,5954(71) $ & \\
12 & $ \mu_h'/\mu_p'       $ &
$-0.761\, 786 \, 1313(33)$  & \\
13 & $ \mu_n/\mu_p'        $ &
$ -0.684\, 996 \, 94(16) $   & \\ \hline
& & & \\[-3mm]
14 & $ m_e/m_\mu           $    &
$4.836\, 332\, 10(15) \times 10^{-3}$  & \\ \hline
& & & \\[-3mm]
15 & $ h                   $   & 
$6.626\, 068\, 76(52) \times 10^{-34}\, {\rm J}\,{\rm s}$    & \\ \hline
& & & \\[-3mm]
16 & $ R                   $   &
$ 8.314 \,472(15)\, {\rm J}\,{\rm mol}^{-1}\,{\rm K}^{-1}$   & \\ \hline
& & & \\[-3mm]
17 & $\rm xu(Cu\:K\alpha_1)$    &$1.002\, 077\, 03(28)\times10^{-13}\,{\rm m }$  & $\lambda({\rm CuK}_{\alpha1}) = 1\,573.400 \rm xu(Cu\:K\alpha_1)$\\
18 & $\rm xu(Mo\:K\alpha_1)$    & $ 1.002\, 099 \,59(53)\times 10^{-13}\,
{\rm m}$ & $\lambda({\rm MoK}_{\alpha1}) = 707.831 \rm xu(Mo\:K\alpha_1)$\\ 19 & $ \rm \AA^*           $    &
$ 1.000\, 015\, 01(90) \times 10^{-10}\, {\rm m}$     & 
$\lambda({\rm WK}_{\alpha1}) = 0.2090100 \rm \AA^* $\\ \hline
20 & $ \rm d_{220}(ILL)    $    &not given  & \\
21 & $ \rm d_{220}(N)      $    &not given   & \\
22 & $ \rm d_{220}(W17)    $    &not given   & \\
23 & $ \rm d_{220}(W04)    $    &not given   & \\
24 & $ \rm d_{220}(W4.2a)  $    & not given  & \\
25 & $ \rm d_{220}(MO^*4)  $    &not given   & \\
26 & $ \rm d_{220}(SH1)    $    &not given   & \\
27 & $ \rm d_{220}         $    & $1.920\,155\,845(56)\times 10^{-10}\,{\rm
 m} $  & \\ \hline
 & & & \\[-3mm]
28 & $ R_p                 $    &$ 0.907(32)\times 10^{-15}\, {\rm m}$  & From \cite{MohrTaylor}, page 440\\
29 & $ R_d                 $    & $ 2.153(14)\times 10^{-15}\, {\rm m}$  &From \cite{MohrTaylor}, page 440 \\  
\hline
\end{tabular}
}
\end{center}

%% file: deltable.tex
\begin{center}
{\small 
\begin{tabular}{|r|l|c|c|}
\hline 
 & & & \\[-3mm]
& NIST Symbol &  NIST/CODATA:1998 Value & Comments\\ 
\hline
   & & & \\[-4mm]

30 & $ {{\delta }_H}(1,0,1) $ & not given  & \\
31 & $ {{\delta }_H}(2,0,1) $ &  not given  & \\
32 & $ {{\delta }_H}(3,0,1) $ &  not given  & \\
33 & $ {{\delta }_H}(4,0,1) $ &   not given & \\
34 & $ {{\delta }_H}(6,0,1) $ &   not given & \\
35 & $ {{\delta }_H}(8,0,1) $ &   not given & \\
36 & $ {{\delta }_H}(2,1,1) $ &   not given & \\
37 & $ {{\delta }_H}(4,1,1) $ &   not given & \\
38 & $ {{\delta }_H}(2,1,3) $ &   not given & \\
39 & $ {{\delta }_H}(4,1,3) $ &   not given & \\
40 & $ {{\delta }_H}(8,2,3) $ &   not given & \\
41 & $ {{\delta }_H}(12,2,3) $ &   not given & \\
42 & $ {{\delta }_H}(4,2,5) $ &   not given & \\
43 & $ {{\delta }_H}(6,2,5) $ &   not given & \\
44 & $ {{\delta }_H}(8,2,5) $ &   not given & \\
45 & $ {{\delta }_H}(12,2,5) $ &   not given & \\                        46 & $ {{\delta }_D}(1,0,1) $ &  not given  & \\                         47 & $ {{\delta }_D}(2,0,1) $ &   not given & \\  
48 & $ {{\delta }_D}(4,0,1) $ &  not given & \\
49 & $ {{\delta }_D}(8,0,1) $ &   not given & \\  
50 & $ {{\delta }_D}(8,2,3) $ &   not given & \\
51 & $ {{\delta }_D}(12,2,3) $ &   not given & \\
52 & $ {{\delta }_D}(4,2,5) $ &   not given & \\
53 & $ {{\delta }_D}(8,2,5) $ &   not given & \\
54 & $ {{\delta }_D}(12,2,5) $ &  not given & \\
55 & $ \delta_e $ &$0.1(1.1)\times10^{-12}$  & From \cite{MohrTaylor}, page
457\\
56 & $ \delta_{Mu} $ &   not given & \\
57 & $ \delta_\mu $ &$0.0(6.4)\times10^{-10}$  &From \cite{MohrTaylor}, page
457 \\ 
\hline
\end{tabular}
}
\end{center}

%% file: table1-1.tex
\begin{center}
{
\begin{tabular}{|r|l|l|l|c|}
\hline 
& & & &\\[-3mm]
& FPC   & ~~~~~~~ NIST:1998 Value   & ~~~~~ IHEP:2003 Value &  \\ 
& Symbol&~~~~ NIST true correlator & NIST published correlator& {\boldmath $\Delta$}\\ 
\hline
 & & & &\\[-2mm]
 1& $R_{\infty}$ & $\phantom{-}1.097\,373\,156\,8549(83)\times 10^{7}$&
                   $\phantom{-}1.097\,373\,156\,854\,7(83)\times10^{7}$ &
                   0.0153 \\ 
                   
 2& $ A_r(e) $ & $ \phantom{-}5.485\,799\,110(12)\times 10^{-4} $ & 
                 $ \phantom{-}5.485\,799\,1097(116)\times 10^{-4} $ &  
 0.0171 \\ 
 
 3& $ A_r(p) $ & $ \phantom{-}1.007\,276\,466\,88(13) $ & 
                 $ \phantom{-}1.007\,276\,466\,883(132) $ & 
$-0.0153$\\ 

4&$ A_r(n) $ & $ \phantom{-}1.008\,664\,915\,78(55) $ & 
               $ \phantom{-}1.008\,664\,915\,784(547) $ & 
$ 0.00501 $ \\ 

5&$ A_r(d) $ & $ \phantom{-}2.013\,553\,212\,71(35) $ & 
               $ \phantom{-}2.013\,553\,212\,706(344) $ & 
$ 0.00833 $ \\ 

6&$ A_r(h) $ & $ \phantom{-}3.014\,932\,234\,69(86) $ & 
               $ \phantom{-}3.014\,932\,234\,691(860) $ & 
$ -0.001 $ \\ 

7&$ A_r(\alpha) $ & $ \phantom{-}4.001\,506\,1747(10) $ & 
                    $ \phantom{-}4.001\,506\,174\,69(100) $ &
$ 0.00456 $\\ 

8&$ \alpha $ & $ \phantom{-}7.297\,352\,533(27)\times 10^{-3} $ & 
               $ \phantom{-}7.297\,352\,5335(265)\times 10^{-3} $ &
$ -0.0132 $\\ 

9&$ \mu_{e^{-}}/\mu_p $ & $ -6.582\,106\,875(66)\times 10^{2} $ & 
                          $ -6.582\,106\,8753(659)\times 10^{2} $ & 
$ 0.00375 $\\ 

10&$ \mu_d/\mu_e^{-} $ & $ -4.664\,345\,537(50)\times 10^{-4} $ & 
                        $ -4.664\,345\,5371(500)\times 10^{-4} $ & 
$ 0.0009 $ \\ 

11&$ \mu_{e^{-}}/\mu_p' $ & $ -6.582\,275\,954(71)\times 10^{2} $ & 
                            $ -6.582\,275\,9549(717)\times 10^{2} $ &
$ 0.00938 $ \\ 

12&$ \mu_h'/\mu_p' $ & $ -7.617\,861\,313(33)\times 10^{-1} $ & 
                       $ -7.617\,861\,3130(330)\times 10^{-1} $ & 
$ 6.64\times{10}^{-9} $ \\ 

13&$ \mu_n/\mu_p' $ & $ -6.849\,9694(16)\times 10^{-1} $ & 
                      $ -6.849\,969\,40(160)\times 10^{-1} $ & 
$ 6.16\times{10}^{-12} $ \\ 

14&$ m_e/m_\mu $ & $ \phantom{-}4.836\,332\,10(15)\times 10^{-3} $ & 
                   $ \phantom{-}4.836\,332\,107(144)\times 10^{-3} $ &
{$ -0.0360 $} \\ 

15&$ h $ & $ \phantom{-}6.626\,068\,76(52)\times 10^{-34} $ & 
           $ \phantom{-}6.626\,068\,756(522)\times 10^{-34} $ &
$ 0.00530 $ \\ 

16&$ R $ & $ \phantom{-}8.314\,472(15) $ & 
           $ \phantom{-}8.314\,4724(147) $ &
$ -0.0214 $ \\ 

17&$\rm xu(Cu\:K\alpha_1)$& $\phantom{-}1.002\,077\,03(28)\times 10^{-13}$& 
                            $\phantom{-}1.002\,077\,021(287)\times 10^{-13}$& $ 0.0212 $ \\ 
                            
18&$\rm xu(Mo\:K\alpha_1)$& $ \phantom{-}1.002\,099\,59(53)\times 10^{-13}$&                            
                            $ \phantom{-}1.002\,099\,593(516)\times 10^{-13}$& $ -0.00461 $ \\ 
                            
19&$ \rm \AA^* $ & $ \phantom{-}1.000\,015\,01(90)\times 10^{-10} $ & 
                   $ \phantom{-}1.000\,015\,010(901)\times 10^{-10} $ & 
$ -0.000017 $ \\ 

20&$ \rm d_{220}(ILL) $ &  ~~~~~~~ not given  & 
                      $ \phantom{-}1.920\,155\,8160(558)\times 10^{-10}$&
--- --- ---\\ 
21&$ \rm d_{220}(N) $ &  ~~~~~~~ not given  & 
                       $ \phantom{-}1.920\,155\,8191(508)\times 10^{-10}$& --- --- ---\\ 
                       
22&$ \rm d_{220}(W17) $ &  ~~~~~~~ not given  & 
                         $ \phantom{-}1.920\,155\,8411(484)\times 10^{-10}$& --- --- ---\\ 
                         
23&$ \rm d_{220}(W04) $ &  ~~~~~~~ not given  & 
                         $ \phantom{-}1.920\,155\,8103(520)\times 10^{-10}$& --- --- ---\\ 
                         
24&$ \rm d_{220}(W4.2a) $ &  ~~~~~~~ not given  & 
                         $ \phantom{-}1.920\,155\,7995(496)\times 10^{-10}$& --- --- ---\\ 
                         
25&$ \rm d_{220}(MO^*4) $ &  ~~~~~~~ not given  & 
                           $ \phantom{-}1.920\,155\,6075(439)\times 10^{-10}$& --- --- ---\\ 
                           
26&$ \rm d_{220}(SH1) $ & ~~~~~~~ not given  & 
                         $ \phantom{-}1.920\,155\,7624(463)\times 10^{-10}$& --- --- ---\\ 
                         
27&$ \rm d_{220} $ & $ \phantom{-}1.920\,155\,845(56)\times 10^{-10} $ &
                     $ \phantom{-}1.920\,155\,8391(561)\times 10^{-10} $& $ 0.0749 $ \\ 
                     
28 & $ R_p       $ &$ \phantom{-}0.907(32)\times 10^{-15}$&
                    $ \phantom{-}0.9066(329)\times 10^{-16} $ &
                    --- --- ---   \\
                    
29 & $ R_d       $ &$ \phantom{-}2.153(14)\times 10^{-15} $&
                   $\phantom{-}2.1528(137)\times 10^{-15}$
                     &--- --- ---   \\
\hline \hline
 & & & &\\[-3mm]
 & & {\boldmath $\chi^2/ndf = 0.90$} \cite{MohrTaylor}, page 446 & ~~~~~~~~{\boldmath $\chi^2/ndf = 0.90$} & \\[1mm]
 \hline
 \end{tabular}
 }
\end{center}

%% file: table1-3.tex
\begin{center}
{\small 
\begin{tabular}{|r|l|l|l|}
\hline
 & & & \\[-3mm]
& FPC   & NIST:1998 input value   & ~~~~~ IHEP:2003 value \\
& Symbol&  & NIST published correlator\\
\hline
 & & & \\[-3mm]
30&$ {{\delta }_H}(1,0,1) $ & $ 0.0000(9.0000)\times 10^{4} $ & 
$ \phantom{-}0.009(8.974)\times 10^{4} $ \\ 
31&$ {{\delta }_H}(2,0,1) $ & $ 0.0000(1.1000)\times 10^{4} $ & 
$ \phantom{-}0.007(1.097)\times 10^{4} $ \\ 
32&$ {{\delta }_H}(3,0,1) $ & $ 0.000(3.300)\times 10^{3} $ & 
$ \phantom{-}0.003(3.291)\times 10^{3} $ \\ 
33&$ {{\delta }_H}(4,0,1) $ & $ 0.000(1.400)\times 10^{3} $ & 
$ \phantom{-}0.009(1.396)\times 10^{3} $ \\ 
34&$ {{\delta }_H}(6,0,1) $ & $ 0.00(4.20)\times 10^{2} $ & 
$ \phantom{-}0.00(4.19)\times 10^{2} $ \\ 
35&$ {{\delta }_H}(8,0,1) $ & $ 0.00(1.80)\times 10^{2} $ & 
$ \phantom{-}0.00(1.79)\times 10^{2} $ \\ 
36&$ {{\delta }_H}(2,1,1) $ & $ 0.000(1.100)\times 10^{3} $ & 
$ -0.04(1.09)\times 10^{3} $ \\ 
37&$ {{\delta }_H}(4,1,1) $ & $ 0.00(1.40)\times 10^{2} $ & 
$ -0.05(1.39)\times 10^{2} $ \\ 
38&$ {{\delta }_H}(2,1,3) $ & $ 0.000(1.100)\times 10^{3} $ & 
$ \phantom{-}0.01(1.10)\times 10^{3} $ \\ 
39&$ {{\delta }_H}(4,1,3) $ & $ 0.00(1.40)\times 10^{2} $ & 
{$ \phantom{-} 0.02(1.39)\times 10^{2} $ }\\ 
40&$ {{\delta }_H}(8,2,3) $ & $ 0.0(1.7)\times 10^{1} $ & 
$ -0.00(1.70)\times 10^{1} $ \\ 
41&$ {{\delta }_H}(12,2,3) $ & $ 0.0(5.0) $ & $ -0.007(5.000) $ \\ 
42&$ {{\delta }_H}(4,2,5) $ & $ 0.00(1.40)\times 10^{2} $ &
{ \phantom{-}$0.01(1.40)\times 10^{2} $} \\ 
43&$ {{\delta }_H}(6,2,5) $ & $ 0.0(4.0)\times 10^{1} $ & 
$\phantom{-} 0.03(4.00)\times 10^{1} $ \\ 
44&$ {{\delta }_H}(8,2,5) $ & $ 0.0(1.7)\times 10^{1} $ & 
{ \phantom{-}$0.10(1.70)\times 10^{1} $} \\ 
45&$ {{\delta }_H}(12,2,5) $ & $ 0.0(5.0) $ & 
$ \phantom{-}0.04(5.00) $ \\ 
46&$ {{\delta }_D}(1,0,1) $ & $ 0.0000(8.9000)\times 10^{4} $ & 
$ \phantom{-}0.005(8.877)\times 10^{4} $ \\ 
47&$ {{\delta }_D}(2,0,1) $ & $ 0.0000(1.1000)\times 10^{4} $ & 
$ \phantom{-}0.001(1.097)\times 10^{4} $ \\ 
48&$ {{\delta }_D}(4,0,1) $ & $ 0.000(1.400)\times 10^{3} $ & 
$ \phantom{-}0.001(1.396)\times 10^{3} $ \\ 
49&$ {{\delta }_D}(8,0,1) $ & $ 0.00(1.70)\times 10^{2} $ & 
$ \phantom{-}0.00(1.70)\times 10^{2} $ \\ 
50&$ {{\delta }_D}(8,2,3) $ & $ 0.0(1.1)\times 10^{1} $ & 
$ -0.00(1.10)\times 10^{1} $ \\ 
51&$ {{\delta }_D}(12,2,3) $ & $ 0.0(3.4) $ & $ -0.00(3.40) $ \\ 
52&$ {{\delta }_D}(4,2,5) $ & $ 0.0(9.2)\times 10^{1} $ & 
$ \phantom{-}0.06(9.20)\times 10^{1} $ \\ 
53&$ {{\delta }_D}(8,2,5) $ & $ 0.0(1.1)\times 10^{1} $ & 
$ \phantom{-}0.008(1.100)\times 10^{1} $ \\ 
54&$ {{\delta }_D}(12,2,5) $ & $ 0.0(3.4) $ & 
$ \phantom{-}0.02(3.40) $ \\ 
55&$ \delta_e $ & $ 0.00(1.1)\times 10^{-12} $ & 
{ $ \phantom{-}0.01(1.10)\times 10^{-12} $} \\ 
56&$ \delta_{Mu} $ & $ 0.00(1.20)\times 10^{2} $ & 
$ \phantom{-}0.01(1.17)\times 10^{2} $ \\ 
57&$ \delta_\mu $ & $ 0.0(6.4)\times 10^{-10} $ & 
$ \phantom{-}0.00(6.40)\times 10^{-10} $ \\
\hline \hline
 & & & \\[-3mm]
 & &  & ~~~~~~~~{\boldmath \,$\chi^2/ndf = 0.90$}  \\[1mm]
 \hline 
 \end{tabular}
}
\end{center}

%% file: table21-1.tex
\begin{center}
{\footnotesize
\begin{tabular}{|r|l|l|l|c|}
\hline 
& & & &\\[-3mm]
& FPC   & ~~~~~ NIST:1998 Value   & ~~~~~ IHEP:2003 Value & \\ 
& Symbol& NIST true correlator & Identity matrix correlator&{\boldmath $\Delta$} \\ 
\hline
 & & & &\\[-2mm]
 1& $ R_{\infty} $ & $ \phantom{-}1.097\,373\,156\,8549(83)\times 10^{7} $  
 & $\phantom{-}1.097\,373\,156\,8545(103)\times 10^{7} $  
 &  $ 0.0314 $ \\  
2& $ A_r(e) $ & $ \phantom{-}5.485\,799\,110(12)\times 10^{-4} $ & 
 $ \phantom{-}5.485\,799\,1099(116)\times 10^{-4} $  
 &   $ 0.00317 $ \\ 
 3& $ A_r(p) $ & $ \phantom{-}1.007\,276\,466\,88(13) $ & $ \phantom{-}1.007\,276\,466\,883(132) $ 
 &   $ -0.0138 $ \\ 
4&$ A_r(n) $ & $ \phantom{-}1.008\,664\,915\,78(55) $ & $ \phantom{-}1.008\,664\,915\,774(556) $  
&  $ 0.00731 $ \\ 
5&$ A_r(d) $ & $ \phantom{-}2.013\,553\,212\,71(35) $ & $ \phantom{-}2.013\,553\,212\,688(360) $  
& $   0.0439 $  \\ 
6&$ A_r(h) $ & $ \phantom{-}3.014\,932\,234\,69(86) $ & $ \phantom{-}3.014\,932\,234\,690(860) $  
&  $ 0.000402 $  \\ 
7&$ A_r(\alpha) $ & $ \phantom{-}4.001\,506\,1747(10) $ & $ \phantom{-} 4.001\,506\,17469(100) $  & $ 0.00459 $\\ 
8&$ \alpha $ & $ \phantom{-}7.297\,352\,533(27)\times 10^{-3} $ & $ \phantom{-} 7.297\,352\,5349(266)\times 10^{-3} $ 
&{ $ -0.0500 $  }\\ 
9&$ \mu_{e^{-}}/\mu_p $ & $ -6.582\,106\,875(66)\times 10^{2} $ & $ -6.582\,106\,8754(659)\times 10^{2} $  
& $ 0.00432 $ \\ 
10&$ \mu_d/\mu_e^{-} $ & $ -4.664\,345\,537(50)\times 10^{-4} $ & $ -4.664\,345\,5371(500)\times 10^{-4} $  
&  $ 0.000901 $  \\ 
11&$ \mu_{e^{-}}/\mu_p' $ & $ -6.582\,275\,954(71)\times 10^{2} $ & $ -6.582\,275\,9550(717)\times 10^{2} $  
&$ 0.00973 $  \\ 
12&$ \mu_h'/\mu_p' $ & $ -7.617\,861\,313(33)\times 10^{-1} $ & $ -7.617\,861\,3130(330)\times 10^{-1} $  & $ 6.64 \times {10}^{-9} $  \\ 
13&$ \mu_n/\mu_p' $ & $ -6.849\,9694(16)\times 10^{-1} $ & $ -6.849\,96940(160)\times 10^{-1} $  & $ 6.16 \times {10}^{-12} $ \\ 
14&$ m_e/m_\mu $ & $ \phantom{-}4.836\,332\,10(15)\times 10^{-3} $ & $ \phantom{-} 4.836\,332\,103(142)\times 10^{-3} $  
&{ $ -0.0124 $ } \\ 
15&$ h $ & $ \phantom{-}6.626\,068\,76(52)\times 10^{-34} $ & $ \phantom{-}6.626\,068\,756(522)\times 10^{-34} $  
& $ 0.00523 $  \\ 
16&$ R $ & $ \phantom{-}8.314\,472(15) $ & $ \phantom{-}8.314\,4724(147) $  
&$ -0.0214 $ \\ 
17&$ \rm xu(Cu\:K\alpha_1) $ & $ \phantom{-}1.002\,077\,03(28)\times 10^{-13} $ 
& $ \phantom{-}1.002\,077\,018(288)\times 10^{-13} $  & $ 0.0306 $  \\ 
18&$ \rm xu(Mo\:K\alpha_1) $ & $ \phantom{-}1.002\,099\,59(53)\times 10^{-13} $ 
& $ \phantom{-} 1.002\,099\,596(516)\times 10^{-13} $  &  $ -0.00803 $  \\ 
19&$ \rm \AA^* $ & $ \phantom{-}1.000\,015\,01(90)\times 10^{-10} $ & 
$ \phantom{-}1.000\,015\,013(901)\times 10^{-10} $  &  $ -0.00200 $  \\ 
20&$ \rm d_{220}(ILL) $ &  ~~~~~~~ not given  & $ \phantom{-}1.920\,155\,8093(517)\times 10^{-10} $ &--- --- ---\\ 
21&$ \rm d_{220}(N) $ &  ~~~~~~~ not given  & $ \phantom{-}1.920\,155\,8239(635)\times 10^{-10} $ &--- --- ---\\ 
22&$ \rm d_{220}(W17) $ &  ~~~~~~~ not given  & $ \phantom{-}1.920\,155\,8380(545)\times 10^{-10} $ &--- --- ---\\ 
23&$ \rm d_{220}(W04) $ &  ~~~~~~~ not given  & $ \phantom{-}1.920\,155\,8102(460)\times 10^{-10} $ &--- --- ---\\ 
24&$ \rm d_{220}(W4.2a) $ &  ~~~~~~~ not given  & $ \phantom{-}1.920\,155\,7910(563)\times 10^{-10} $ &--- --- ---\\ 
25&$ \rm d_{220}(MO^*4) $ &  ~~~~~~~ not given  & $ \phantom{-}1.920\,155\,5957(457)\times 10^{-10} $ &--- --- ---\\ 
26&$ \rm d_{220}(SH1) $ & ~~~~~~~ not given  & $ \phantom{-}1.920\,155\,7631(509)\times 10^{-10} $ &--- --- ---\\ 
27&$ \rm d_{220} $ & $ \phantom{-}1.920\,155\,845(56)\times 10^{-10} $ 
& $ \phantom{-}1.920\,155\,8390(506)\times 10^{-10} $  &  $ 0.0800 $ \\
28 & $ R_p       $ &$ \phantom{-}0.907(32)\times 10^{-15}$&
                    $ \phantom{-}0.9087(261)\times 10^{-15} $ &
                     --- --- ---  \\
29 & $ R_d       $ &$ \phantom{-}2.153(14)\times 10^{-15} $&
                    $ \phantom{-}2.1519(115)\times 10^{-15} $ & 
                    --- --- ---  \\
\hline \hline
 & & & &\\[-3mm]
 & & {\boldmath $\chi^2/ndf = 0.90$} \cite{MohrTaylor}, page 446 & ~~~~~~~~{\boldmath $\chi^2/ndf = 0.84$} & \\[1mm]
 \hline
 \end{tabular}
 }
\end{center}

%% file: table22-3.tex
\begin{center}
{
\begin{tabular}{|r|l|l|l|}
\hline
 & & & \\[-3mm]
& FPC   & NIST:1998 input value   & ~~~~~ IHEP:2003 value \\
& Symbol& & Identity matrix correlator\\
\hline
 & & & \\[-3mm]
30&$ {{\delta }_H}(1,0,1) $ & $ 0.0000(9.0000)\times 10^{4} $ & 
$ -0.06(6.40)\times 10^{4} $  \\ 
31&$ {{\delta }_H}(2,0,1) $ & $ 0.0000(1.1000)\times 10^{4} $ & 
$ -0.07(7.96)\times 10^{3} $ \\ 
32&$ {{\delta }_H}(3,0,1) $ & $ 0.000(3.300)\times 10^{3} $ & 
$ \phantom{-}0.01(3.29)\times 10^{3} $ \\ 
33&$ {{\delta }_H}(4,0,1) $ & $ 0.000(1.400)\times 10^{3} $ & 
$ \phantom{-}0.01(1.39)\times 10^{3} $  \\ 
34&$ {{\delta }_H}(6,0,1) $ & $ 0.00(4.20)\times 10^{2} $ & 
$ \phantom{-}0.02(4.20)\times 10^{2} $  \\ 
35&$ {{\delta }_H}(8,0,1) $ & $ 0.00(1.80)\times 10^{2} $ & 
$ -0.01(1.80)\times 10^{2} $  \\ 
36&$ {{\delta }_H}(2,1,1) $ & $ 0.000(1.100)\times 10^{3} $ & 
$ -0.05(1.10)\times 10^{3} $  \\ 
37&$ {{\delta }_H}(4,1,1) $ & $ 0.00(1.40)\times 10^{2} $ & 
$ \phantom{-}0.01(1.40)\times 10^{2} $  \\ 
38&$ {{\delta }_H}(2,1,3) $ & $ 0.000(1.100)\times 10^{3} $ & 
$ \phantom{-}0.03(1.10)\times 10^{3} $  \\ 
39&$ {{\delta }_H}(4,1,3) $ & $ 0.00(1.40)\times 10^{2} $ & 
$ -0.02(1.40)\times 10^{2} $ \\ 
40&$ {{\delta }_H}(8,2,3) $ & $ 0.0(1.7)\times 10^{1} $ & 
$\phantom{-}0.00(1.70)\times 10^{1} $  \\ 
41&$ {{\delta }_H}(12,2,3) $ & $ 0.0(5.0) $ & $ -0.00(5.00) $  \\ 
42&$ {{\delta }_H}(4,2,5) $ & $ 0.00(1.40)\times 10^{2} $ &
$ \phantom{-}0.005(1.400)\times 10^{2} $ \\ 
43&$ {{\delta }_H}(6,2,5) $ & $ 0.0(4.0)\times 10^{1} $ & 
$ -0.009(4.000)\times 10^{1} $ \\ 
44&$ {{\delta }_H}(8,2,5) $ & $ 0.0(1.7)\times 10^{1} $ & 
$ \phantom{-}0.000(1.700)\times 10^{1} $  \\ 
45&$ {{\delta }_H}(12,2,5) $ & $ 0.0(5.0) $ & 
 $ -0.00(5.00) $ \\ 
46&$ {{\delta }_D}(1,0,1) $ & $ 0.0000(8.9000)\times 10^{4} $ & 
$ \phantom{-}0.05(6.50)\times 10^{4} $  \\ 
47&$ {{\delta }_D}(2,0,1) $ & $ 0.0000(1.1000)\times 10^{4} $ & 
$ -0.04(8.07)\times 10^{3}$ \\ 
48&$ {{\delta }_D}(4,0,1) $ & $ 0.000(1.400)\times 10^{3} $ & 
$ -0.08(1.40)\times 10^{3} $ \\ 
49&$ {{\delta }_D}(8,0,1) $ & $ 0.00(1.70)\times 10^{2} $ & 
$ \phantom{-}0.02(1.70)\times 10^{2} $  \\ 
50&$ {{\delta }_D}(8,2,3) $ & $ 0.0(1.1)\times 10^{1} $ & 
$ -0.00(1.10)\times 10^{1} $  \\ 
51&$ {{\delta }_D}(12,2,3) $ & $ 0.0(3.4) $ & $ -0.00(3.40) $ \\ 
52&$ {{\delta }_D}(4,2,5) $ & $ 0.0(9.2)\times 10^{1} $ & 
$ \phantom{-}0.00(9.20)\times 10^{1} $  \\ 
53&$ {{\delta }_D}(8,2,5) $ & $ 0.0(1.1)\times 10^{1} $ & 
$ \phantom{-}0.00(1.10)\times 10^{1} $  \\ 
54&$ {{\delta }_D}(12,2,5) $ & $ 0.0(3.4) $ & $ -0.00(3.40) $  \\ 
55&$ \delta_e $ & $ 0.0(1.1)\times 10^{-12} $ & 
$ \phantom{-}0.08(1.10)\times 10^{-12}$ \\ 
56&$ \delta_{Mu} $ & $ 0.00(1.20)\times 10^{2} $ & 
$ \phantom{-}0.01(1.17)\times 10^{2} $ \\ 
57&$ \delta_\mu $ & $ 0.0(6.4)\times 10^{-10} $ & 
$ \phantom{-}0.00(6.40)\times 10^{-10} $ \\
\hline \hline
 & & & \\[-3mm]
 & & & ~~~~~~~~{\boldmath $\chi^2/ndf = 0.84$}  \\[1mm]
 \hline 
 \end{tabular}
}
\end{center}

%% file: table3-1-1.tex
\begin{center}
{
\begin{tabular}{|r|l|l|l|c|}
\hline 
& & & &\\[-3mm]
& FPC   & ~~~~~ IHEP:2003 value   & ~~~~~ IHEP:2003 Value &  \\ 
& Symbol&NIST published correlator &Identity matrix correlator &{\boldmath
$\Delta$}\\ \hline
 & & & &\\[-2mm] 
$ 1 $ & $ R_{\infty} $ & $ \phantom{-}1.097\:373\:156\:855(8)\times 10^{7} $ & $ \phantom{-}1.097\:373\:156\:854(10)\times 10^{7} $ & $ \phantom{-}0.018 $ \\ 
$ 2 $ & $ R_p $ & $ \phantom{-}9.066(329)\times 10^{-16} $ & 
$ \phantom{-}9.087(261)\times 10^{-16} $ & $ -0.048 $ \\ 
$ 3 $ & $ R_d $ & $ \phantom{-}2.1528(137)\times 10^{-15} $ & 
$ \phantom{-}2.1519(115)\times 10^{-15} $ & $ \phantom{-}0.048 $ \\ 
$ 4 $ & $ A_r(e) $ & $ \phantom{-}5.485\:799\:1097(116)\times 10^{-4} $ & $ \phantom{-}5.485\:799\:1099(116)\times 10^{-4} $ & $ -0.014 $ \\ 
$ 5 $ & $ A_r(p) $ & $ \phantom{-}1.007\:276\:466\:883(132) $ & $ \phantom{-}1.007\:276\:466\:883(132) $ & $ \phantom{-}0.0015 $ \\ 
$ 6 $ & $ A_r(n) $ & $ \phantom{-}1.008\:664\:915\:784(547) $ & $ \phantom{-}1.008\:664\:915\:774(566) $ & $ \phantom{-}0.012 $ \\ 
$ 7 $ & $ A_r(d) $ & $ \phantom{-}2.013\:553\:212\:706(344) $ & $ \phantom{-}2.013\:553\:212\:688(360) $ & $ \phantom{-}0.036 $ \\ 
$ 8 $ & $ A_r(h) $ & $ \phantom{-}3.014\:932\:234\:691(860) $ & $ \phantom{-}3.014\:932\:234\:690(860) $ & $ \phantom{-}0.0014 $ \\ 
$ 9 $ & $ A_r(\alpha) $ & $ \phantom{-}4.001\:506\:174\:69(100) $ & $ \phantom{-}4.001\:506\:174\:69(100) $ & $ \phantom{-}0.000\:33 $ \\ 
$ 10 $ & $ \alpha $ & $ \phantom{-}7.297\:352\:5335(265)\times 10^{-3} $ & $ \phantom{-}7.297\:352\:5349(266)\times 10^{-3} $ & $ -0.037 $ \\ 
$ 11 $ & $ \mu_{e^{-}}/\mu_p $ & $ -6.582\:106\:8753(659)\times 10^{2} $ & $ -6.582\:106\:8754(659)\times 10^{2} $ & $ \phantom{-}0.0058 $ \\ 
$ 12 $ & $ \mu_d/\mu_e^{-} $ & $ -4.664\:345\:5371(500)\times 10^{-4} $ & $ -4.664\:345\:5371(500)\times 10^{-4} $ & $ \phantom{-}0 $ \\ 
$ 13 $ & $ \mu_{e^{-}}/\mu_p' $ & $ -6.582\:275\:9549(717)\times 10^{2} $ & $ -6.582\:275\:9550(717)\times 10^{2} $ & $ \phantom{-}0.0035 $ \\ 
$ 14 $ & $ \mu_h'/\mu_p' $ & $ -7.617\:861\:3130(330)\times 10^{-1} $ & $ -7.617\:861\:3130(330)\times 10^{-1} $ & $ \phantom{-}0 $ \\ 
$ 15 $ & $ \mu_n/\mu_p' $ & $ -6.849\:969\:40(160)\times 10^{-1} $ & $ -6.849\:969\:40(160)\times 10^{-1} $ & $ \phantom{-}0 $ \\ 
$ 16 $ & $ m_e/m_\mu $ & $ \phantom{-}4.836\:332\:107(144)\times 10^{-3} $ & $ \phantom{-}4.836\:332\:103(142)\times 10^{-3} $ & $ \phantom{-}0.024 $ \\ 
$ 17 $ & $ h $ & $ \phantom{-}6.626\:068\:756(522)\times 10^{-34} $ & $ \phantom{-}6.626\:068\:756(522)\times 10^{-34} $ & $ -0.000\:64 $ \\ 
$ 18 $ & $ R $ & $ \phantom{-}8.314\:4724(147) $ & $ \phantom{-}8.314\:4724(147) $ & $ \phantom{-}0 $ \\ 
$ 19 $ & $ \rm xu(Cu\:K\alpha_1) $ & $ \phantom{-}1.002\:077\:021(287)\times 10^{-13} $ & $ \phantom{-}1.002\:077\:018(288)\times 10^{-13} $ & $ \phantom{-}0.0093 $ \\ 
$ 20 $ & $ \rm xu(Mo\:K\alpha_1) $ & $ \phantom{-}1.002\:099\:593(516)\times 10^{-13} $ & $ \phantom{-}1.002\:099\:596(516)\times 10^{-13} $ & $ -0.0035 $ \\ 
$ 21 $ & $ \rm \AA^* $ & $ \phantom{-}1.000\:015\:010(901)\times 10^{-10} $ & $ \phantom{-}1.000\:015\:013(901)\times 10^{-10} $ & $ -0.0020 $ \\ 
$ 22 $ & $ \rm d_{220}(ILL) $ & $ \phantom{-}1.920\:155\:8160(558)\times 10^{-10} $ & $ \phantom{-}1.920\:155\:8093(517)\times 10^{-10} $ & $ \phantom{-}0.087 $ \\ 
$ 23 $ & $ \rm d_{220}(N) $ & $ \phantom{-}1.920\:155\:8191(508)\times 10^{-10} $ & $ \phantom{-}1.920\:155\:8239(635)\times 10^{-10} $ & $ -0.060 $ \\ 
$ 24 $ & $ \rm d_{220}(W17) $ & $ \phantom{-}1.920\:155\:8411(484)\times 10^{-10} $ & $ \phantom{-}1.920\:155\:8380(545)\times 10^{-10} $ & $ \phantom{-}0.043 $ \\ 
$ 25 $ & $ \rm d_{220}(W04) $ & $ \phantom{-}1.920\:155\:8103(520)\times 10^{-10} $ & $ \phantom{-}1.920\:155\:8102(460)\times 10^{-10} $ & $ \phantom{-}0.0015 $ \\ 
$ 26 $ & $ \rm d_{220}(W4.2a) $ & $ \phantom{-}1.920\:155\:7995(496)\times 10^{-10} $ & $ \phantom{-}1.920\:155\:7910(563)\times 10^{-10} $ & $ \phantom{-}0.011 $ \\ 
$ 27 $ & $ \rm d_{220}(MO^*4) $ & $ \phantom{-}1.920\:155\:6075(439)\times 10^{-10} $ & $ \phantom{-}1.920\:155\:5957(457)\times 10^{-10} $ & $ \phantom{-}0.019 $ \\ 
$ 28 $ & $ \rm d_{220}(SH1) $ & $ \phantom{-}1.920\:155\:7624(463)\times 10^{-10} $ & $ \phantom{-}1.920\:155\:7631(509)\times 10^{-10} $ & $ -0.0096 $ \\ 
$ 29 $ & $ \rm d_{220} $ & $ \phantom{-}1.920\:155\:8391(561)\times 10^{-10} $ & $ \phantom{-}1.920\:155\:8390(506)\times 10^{-10} $ & $ \phantom{-}0.0013 $ \\ 

\hline \hline
 & & & &\\[-3mm]
 & & ~~~~~~~~~{\boldmath \,$\chi^2/ndf = 0.90$}  
 & ~~~~~~~~{\boldmath $\chi^2/ndf = 0.84$} & \\[1mm]
 \hline
 \end{tabular}
 }
\end{center}

%% file: table3-2-1.tex
\begin{center}
{
\begin{tabular}{|r|l|c|c|c|}
\hline 
& & & &\\[-3mm]
& FPC   & ~~~~~ IHEP:2003 value   & ~~~~~ IHEP:2003 Value &  \\ 
& Symbol&  NIST published correlator&Identity matrix correlator &{\boldmath
$\Delta$}\\ \hline
 & & & &\\[-2mm]
 
$ 30 $ & $ {{\delta }_H}(1,0,1) $ & $ \phantom{-}0.009(8.974)\times 10^{4} $ & $ -0.062(6.40)\times 10^{4} $ & $ \phantom{-}0.057 $ \\ 
$ 31 $ & $ {{\delta }_H}(2,0,1) $ & $ \phantom{-}0.007(1.0974)\times 10^{4} $ & $ -0.067(7.96)\times 10^{3} $ & $ \phantom{-}0.050 $ \\ 
$ 32 $ & $ {{\delta }_H}(3,0,1) $ & $ \phantom{-}0.003(3.291)\times 10^{3} $ & $ \phantom{-}0.012(3.29)\times 10^{3} $ & $ -0.024 $ \\ 
$ 33 $ & $ {{\delta }_H}(4,0,1) $ & $ \phantom{-}0.009(1.3963)\times 10^{3} $ & $ \phantom{-}0.014(1.39)\times 10^{3} $ & $ -0.069 $ \\ 
$ 34 $ & $ {{\delta }_H}(6,0,1) $ & $ \phantom{-}0(4.19)\times 10^{2} $ & $ \phantom{-}0.02(4.20)\times 10^{2} $ & $ -0.0027 $ \\ 
$ 35 $ & $ {{\delta }_H}(8,0,1) $ & $ \phantom{-}0(1.79)\times 10^{2} $ & $ -0.01(1.80)\times 10^{2} $ & $ \phantom{-}0.0056 $ \\ 
$ 36 $ & $ {{\delta }_H}(2,1,1) $ & $ -0.04(1.09)\times 10^{3} $ & $ -0.05(1.10)\times 10^{3} $ & $ \phantom{-}0.0069 $ \\ 
$ 37 $ & $ {{\delta }_H}(4,1,1) $ & $ -0.05(1.39)\times 10^{2} $ & $ \phantom{-}0.01(1.40)\times 10^{2} $ & $ -0.032 $ \\ 
$ 38 $ & $ {{\delta }_H}(2,1,3) $ & $ \phantom{-}0.01(1.10)\times 10^{3} $ & $ \phantom{-}0.03(1.10)\times 10^{3} $ & $ -0.011 $ \\ 
$ 39 $ & $ {{\delta }_H}(4,1,3) $ & $ \phantom{-}0.02(1.39)\times 10^{2} $ & $ -0.02(1.40)\times 10^{2} $ & $ \phantom{-}0.019 $ \\ 
$ 40 $ & $ {{\delta }_H}(8,2,3) $ & $ -0(1.70)\times 10^{1} $ & $ \phantom{-}0(1.70)\times 10^{1} $ & $ -0.0013 $ \\ 
$ 41 $ & $ {{\delta }_H}(12,2,3) $ & $ -0.007(5.000) $ & $ -0(5.00) $ & $ -0.0060 $ \\ 
$ 42 $ & $ {{\delta }_H}(4,2,5) $ & $ \phantom{-}0.01(1.400)\times 10^{2} $ & $ \phantom{-}0.005(1.400)\times 10^{2} $ & $ \phantom{-}0.0024 $ \\ 
$ 43 $ & $ {{\delta }_H}(6,2,5) $ & $ \phantom{-}0.03(4.00)\times 10^{1} $ & $ -0.009(4.000)\times 10^{1} $ & $ \phantom{-}0.0065 $ \\ 
$ 44 $ & $ {{\delta }_H}(8,2,5) $ & $ \phantom{-}0.01(1.70)\times 10^{1} $ & $ \phantom{-}0(1.70)\times 10^{1} $ & $ \phantom{-}0.0030 $ \\ 
$ 45 $ & $ {{\delta }_H}(12,2,5) $ & $ \phantom{-}0.04(5.00) $ & $ -0(5.00) $ & $ \phantom{-}0.0053 $ \\ 
$ 46 $ & $ {{\delta }_D}(1,0,1) $ & $ \phantom{-}0.0049(8.8767)\times 10^{4} $ & $ \phantom{-}0.046(6.50)\times 10^{4} $ & $ -0.041 $ \\ 
$ 47 $ & $ {{\delta }_D}(2,0,1) $ & $ \phantom{-}0.0011(1.0970)\times 10^{4} $ & $ -0.04(8.07)\times 10^{3} $ & $ \phantom{-}0.0034 $ \\ 
$ 48 $ & $ {{\delta }_D}(4,0,1) $ & $ \phantom{-}0.001(1.396)\times 10^{3} $ & $ -0.08(1.40)\times 10^{3} $ & $ \phantom{-}0.043 $ \\ 
$ 49 $ & $ {{\delta }_D}(8,0,1) $ & $ \phantom{-}0(1.70)\times 10^{2} $ & $ \phantom{-}0.02(1.70)\times 10^{2} $ & $ -0.0076 $ \\ 
$ 50 $ & $ {{\delta }_D}(8,2,3) $ & $ -0(1.10)\times 10^{1} $ & $ \phantom{-}0(1.10)\times 10^{1} $ & $ -0.0012 $ \\ 
$ 51 $ & $ {{\delta }_D}(12,2,3) $ & $ -0(3.40) $ & $ -0(3.40) $ & $ -0.0080 $ \\ 
$ 52 $ & $ {{\delta }_D}(4,2,5) $ & $ \phantom{-}0.06(9.20)\times 10^{1} $ & $ \phantom{-}0(9.20)\times 10^{1} $ & $ \phantom{-}0.0045 $ \\ 
$ 53 $ & $ {{\delta }_D}(8,2,5) $ & $ \phantom{-}0.008(1.100)\times 10^{1} $ & $ \phantom{-}0(1.10)\times 10^{1} $ & $ \phantom{-}0.0039 $ \\ 
$ 54 $ & $ {{\delta }_D}(12,2,5) $ & $ \phantom{-}0.02(3.40) $ & $ -0(3.40) $ & $ \phantom{-}0.0052 $ \\ 
$ 55 $ & $ \delta_e $ & $ \phantom{-}0.010(1.10)\times 10^{-12} $ & $ \phantom{-}0.08(1.10)\times 10^{-12} $ & $ \phantom{-}0.0091 $ \\ 
$ 56 $ & $ \delta_{Mu} $ & $ \phantom{-}0.013(1.17)\times 10^{2} $ & $ \phantom{-}0.012(1.17)\times 10^{2} $ & $ \phantom{-}0.0058 $ \\ 
$ 57 $ & $ \delta_\mu $ & $ \phantom{-}0(6.40)\times 10^{-10} $ & $ \phantom{-}0(6.40)\times 10^{-10} $ & $ \phantom{-}0.000\:20 $ \\ 

\hline \hline
 & & & &\\[-3mm]
 & & {\boldmath \,$\chi^2/ndf = 0.90$} & ~~~~~~~~{\boldmath 
 $\chi^2/ndf = 0.84$}&  \\[1mm]
 \hline 
 \end{tabular}
}
\end{center}

%% file: cor-spec.tex
\begin{center}
\begin{tabular}{c|cccc}
 $          $ & $ R_{\infty} $ & $ \alpha  $ & $ m_e/m_\mu $ & $ h     
 $ \\[1mm] \hline
 & & & & \\[-4mm]
$ R_{\infty} $ & $ 1.00       $ &{\boldmath $ -0.020 $} & 
{\boldmath $0.004 $} & {\boldmath$ -0.000 $} \\ 
$ \alpha     $ & $ -0.0112    $ & $ 1.00    $ & {\boldmath $ -0.233 $} & 
{\boldmath $ 0.002$} \\ 
$ m_e/m_\mu  $ & $ 0.00235    $ & $ -0.236  $ & $ 1.00      $ & 
{\boldmath $ -0.000 $} \\ 
$ h          $ & $ -0.0000195 $ & $ 0.00174 $ & $ -0.000410 $ & $ 1.00   $ \\ 
\end{tabular}
\end{center}